\documentclass[showpacs,preprintnumbers,prd,nofootinbib,floats,amssymb,floatfix]{revtex4}
\usepackage{graphicx}
\usepackage{amsxtra}
\usepackage{hyperref}
\usepackage{amssymb}
\usepackage{amstext}
\usepackage{amsmath}
\usepackage{cleveref}
\usepackage{stackrel}
\usepackage{blkarray}
\begin{document}
\title{ Hamiltonian  analysis for perturbative  $\lambda R$ gravity }
\author{Alberto Escalante}  \email{aescalan@ifuap.buap.mx}
\author{P. Fernando Ocaña-Garc{\'i}a}  \email{pfgarcia@ifuap.buap.mx}
 \affiliation{Instituto de F\'isica, Benem\'erita Universidad Aut\'onoma de Puebla. \\ Apartado Postal J-48 72570, Puebla Pue., M\'exico, }

\begin{abstract}
The Hamiltonian analysis for the linearized  $\lambda R$ gravity around the Minkowski background is performed. The first-class and second-class constraints for arbitrary values of $\lambda$ are presented, and two physical degrees of freedom are reported. In addition, we remove the second-class constraints, and the generalized Dirac brackets are constructed; then, the equivalence between General Relativity and the  $\lambda R$ theory is shown.     
\end{abstract}
 \date{\today}
\preprint{}
\maketitle

\section{Introduction}
It is well-known that including higher-order derivative terms to the Einstein-Hilbert $[EH]$ action improves the behavior in the UV sector of quantum gravity \cite{Stell}. However, the ensuing presence of higher time derivatives leads us to deal with ghost degrees of freedom or the problem of unitarity \cite{OD}. On the other hand, there is an alternative way to add higher-order derivatives to the $EH$ theory. In this regard, Hořava proposed the anisotropic treatment of spacetime, which entailed an entirely new perspective in the search for a consistent theory of quantum gravity \cite{H1,H2}. Unlike the well-known diffeomorphisms invariance of General Relativity $[GR]$, anisotropic spacetime causes Hořava's gravity to be invariant concerning a more restricted group, the so-called foliation-preserving diffeomorphisms. As a result, by employing the Arnowitt-Deser-Misner formalism [$ADM$] \cite{ADM}, a theory containing higher spatial derivatives while keeping time derivatives up to second order is obtained. This theory is power-counting  renormalizable by construction but avoids the ghost problem.\\ 
It is important to mention that this model has been used to account for the  luminal propagation of gravitational waves in agreement with GW170817 and GRB170817A events \cite{Noemi}, and has given rise to a dark energy model \cite{ParkD} that explains naturally the non-interacting nature of the dark
energy sector and that improves the situation of the so-called  \textit{discordance problem} involving the Hubble constant $H_{0}$ and cosmic shear parameter $S_{8}$ \cite{ ParkD2}. Furthermore, within the cosmological context, there are works where inflation was studied. For example, in \cite{cosmo}  by using the Hořava theory the  slow-roll conditions in the Friedmann-Robertson-Walker  background are reported. In fact, it is shown that the gauge invariants for cosmological perturbations are different from those given in $GR$. Moreover, the power spectra and spectrum index of the scalar perturbations in the slow-roll approximations are calculated. By making a direct comparison with $GR$, it is found that the power spectrum and index acquire tiny corrections from the Hořava theory.\\
On the other hand, in \cite{cosmo2} the quantum cosmology for a (1+1)-dimensional Hořava theory is studied. Compared with $GR$, which is a topological theory in two dimensions, the classical two-dimensional Hořava theory is not. In fact, there are propagating degrees of freedom resembling the Jackiw-Teitelboim model, in which a dilatonic degree of freedom is necessary for the dynamics. However, in Hořava's theory, the degree of freedom emerges naturally. In general, it is shown that in $GR$ and  Hořava theory, quantization seems to smooth out the big-bang singularity when the scale factor vanishes while still retaining the classical behavior as the universe becomes asymptotically large.
\\
On the other side, by Horava's requirements, different versions of the theory can be constructed (see \cite{Wang,He} for a comprehensive exposition). Among them is the so-called non-projective version, characterized by the lapse function, which can be a general function of time and space. This version has a limit at large distances that resembles $GR$; this is desirable for any theory that pretends to be a generalization of gravity.   In this respect, Horava's proposal has given rise to valuable discussions on its gauge group and the generic presence of an extra degree of freedom. In fact, it has been argued that regardless of the version, Hořava's theory could present three degrees of freedom, one more than $GR$ \cite{c1,c2}. In the projectable version (the lapse function is restricted to depend only on time) this extra mode is present at all scales \cite{Koba}; however, some works support the consistency of the non-projectable case and have prompted further analysis of it \cite{Das, bellorin, bellorin2,bellorin3}. \\ 
In this paper,  we will focus on the $\lambda R$ model \cite{tesis}, which can be interpreted as a modification of $GR$ \cite{Kiefer} or the truncation of the potential at lowest order in the curvature of the non-projectable Hořava theory with detailed balance, becoming dominant at largest distances (deep IR). The interest in this model is motivated, as commented above, mainly by the alleged existence of an additional degree of freedom with an apparent strong coupling at the extremely low IR,
which was assumed as an inevitable consequence of any model possessing the same foliation-preserving symmetry \cite{c1,c2}. The $\lambda R$ model has provided evidence in favor of the theoretical consistency of Hořava's theory; at the non-perturbative level, it has been shown in \cite{bellorin} through the Hamiltonian analysis that this model consistently describes the dynamics of two physical degrees of freedom, just as in $GR$. Moreover, this model is equivalent to $GR$ in a particular gauge (where $K=0$, the so-called maximal slicing gauge). The condition $K=0$ emerges as a second-class constraint; therefore, the value of $\lambda$ is not relevant, and $GR$ is consistently recovered. \\
Furthermore, in the perturbative sector, there are works focused on determining the number of degrees of freedom by implementing the so-called Hořava's gauge \cite{park1, park2, park3, gong}. However, this type of analysis could lead to incomplete conclusions because not all the constraints present in the theory are correctly identified. In fact, the correct identification and classification of constraints have allowed us to address essential issues in developing and analyzing any gauge theory. For example, in \cite{Hen}, an inconsistency related to the lapse function is reported; however,  in \cite{bellorin}, it is shown that the inconsistency is related to the study of the constraints. In fact, a second-class constraint emerges and restricts another one, the trace of the canonical momenta conjugated to the spatial metric. The preservation in time of this new second-class constraint leads to an equation that allows us to determine the lapse function as a Lagrange multiplier. In this manner, a reliable counting of degrees of freedom is performed, yielding two.
Moreover,  in \cite{Kluson},  it is argued that Horava gravity has a nonvanishing Hamiltonian and does not present one of the aspects related to the so-called “problem of time”. Then,  in \cite{Ted},  all first-class constraints of the theory were analyzed and was confirmed the persistence of the global version of the problem of time. Hence, we can observe that the study of the constraints is mandatory. It is worth noting that first-class constraints are the generators of gauge transformations, and they are used for the identification, for instance,  of observables. On the other hand, second-class constraints allow us to identify the number of Lagrange multipliers that can be found. In addition,  it is well known that the second-class constraints are useful for constructing the Dirac brackets, which are fundamental for the quantization program \cite{13,14}. \\ 
For the reasons explained above, in this work, by considering a perturbative point of view,  we report a detailed canonical analysis of the $\lambda R$ model around a Minkowski background. To this end,  we will consider the 3+1 formalism instead of the linearized $ADM$ formulation; this turns out to be quite convenient since our results can be directly compared with those of linearized gravity reported in the literature \cite{Bar}, where a perturbation around a Minkowski background is considered, and the 3+1 formalism is the standard way for performing the canonical analysis.
\\
The paper is organized as follows. In Section I, the essential tools of $\lambda R$ gravity are presented. In Section II, from the standard linearized action of gravity, we introduce a new set of variables; thus, linearized gravity will be written like Hořava's theory, then the canonical analysis is performed for different values of $\lambda$. The constraints, the Dirac brackets,  and the counting of physical degrees of freedom are reported. In Section V the conclusions are presented.
\section{Linearized $\lambda R$ gravity}
The  $\lambda R$ model in terms of $ADM$ variables  is given by \cite{bellorin}
\begin{equation}\label{full}
    S=\int dt d^{3}x\sqrt{g}N\left(G^{ijkl}K_{ij}K_{kl}+R\right),
\end{equation}
where $N$ is the lapse function, $g_{ij}$ is the spatial metric defined on each spacelike hypersurface, $R$ is the spatial Ricci scalar, $K_{ij}=\frac{1}{2N} \left(\dot{g}_{ij}-2\nabla_{(i}N_{j)}\right)$ is the extrinsic curvature and $G^{ijkl}$ is a generalization of the De Witt metric defined by 
\begin{equation}\label{Wi}
 G^{ijkl}=\frac{1}{2}\left(g^{ik}g^{jl}+g^{il}g^{jk}\right)-\lambda g^{ij}g^{kl}. 
\end{equation}
 The constant $\lambda$ is introduced to establish the separate  compatibility of the kinetic terms with the foliation-preserving diffeomorphisms;  for $\lambda=1$ $EH$ action is recovered. This theory considers a preferred foliation, and the invariance diffeomorphism group is the one that preserves this structure, given by  
\begin{equation}
    t\rightarrow t'(t),\hspace{2cm}x^{i}\rightarrow x'^{i}(x^{i},t),
\end{equation}
 in coordinates adapted to the foliation. The action (\ref{full}) has been analyzed  at linearized level in \cite{park1, park2} where the perturbation around a Minkowski background in the $ADM$ formalism was developed, this is 
 \begin{equation}
     g_{ij}=\delta_{ij}+\epsilon h_{ij},\hspace{1cm}N=1+\epsilon n,\hspace{1cm}N_{i}=\epsilon n_{i},
 \end{equation}
 where $\epsilon$ is an infinitesimal parameter.  However, a complete canonical analysis was not developed; the complete identification of the constraints and the Dirac brackets were not reported. In view of this, we will develop a canonical analysis by working with the 3+1 formalism. In fact, we will use the linearized $EH$ action and a new set of variables. This will allow us to write the action in a new fashion, and the canonical analysis will be done directly.  \\
In this manner, bearing in mind that  $\lambda R$   is a “slight” deviation from $GR$ characterized by the parameter $\lambda$, we will consider the well-known Fierz-Pauli Lagrangian for massless particles of spin two \cite{fierz} and that describes linearized gravity, in its $3+1$ form (see Appendix A)
\begin{equation}\label{FP}
    \begin{split}
        \mathcal{L}_{FP}=&\frac{1}{4}\Dot{h}_{ij}\Dot{h}^{ij}-\Dot{h}^{ij}\partial_{i}h_{0j}-\dot{h}_{j}^{j}\partial_{i}h^{0i}-\frac{1}{4}(\dot{h}_{i}^{i})^{2}-\frac{1}{2}\partial_{i}h_{0j}\partial^{i}h^{0j}+\frac{1}{2}\partial^{i}h^{j0}\partial_{j}h_{i0}+\frac{1}{2}\partial_{i}h_{00}\partial_{j}h^{ij}\\&\hspace{4mm}-\frac{1}{2}\partial_{i}h_{k}^{k}\partial_{j}h^{ij}-\frac{1}{2}\partial_{i}h_{00}\partial^{i}h_{k}^{k}+\frac{1}{4}\partial_{i}h_{j}^{j}\partial^{i}h_{k}^{k}+\frac{1}{2}\partial^{i}h^{jk}\partial_{j}h_{ik}-\frac{1}{4}\partial_{i}h_{jk}\partial^{i}h^{jk},
    \end{split}
\end{equation}
where the perturbation  is given by $g_{\mu\nu}=\eta_{\mu\nu}+h_{\mu\nu}$ with $\eta_{\mu\nu}=\mathrm{diag}(-1,1,1,1)$. By employing an extrinsic curvature type variable given by
\begin{equation}\label{k}
    K_{ij}=\frac{1}{2}\left(\Dot{h}_{ij}-\partial_{i}h_{0j}-\partial_{j}h_{0i}\right),
\end{equation}
we can introduce $\lambda$ into the theory by expressing the Lagrangian  in such a way that it resembles the   $\lambda R$ action (\ref{full}). Thus,  we arrive to  the following new expression  
\begin{equation}\label{Lambda}
    \mathcal{L}=G^{ijkl}K_{ij}K_{kl}-\frac{1}{2}h^{00}R_{ij}^{\hspace{2mm}ij}-\frac{1}{2}h^{ij}\left(R_{ikj}^{\hspace{3.5mm}k}-\frac{1}{2}\delta_{ij}R_{lm}^{\hspace{2mm}lm}\right),
\end{equation}
where 
\begin{equation}
    \begin{split}
 R_{ikj}^{\hspace{3.5mm}k}&=\frac{1}{2}\left(\partial_{k}\partial_{i}h^{k}_{j}-\partial^{k}\partial_{k}h_{ij}-\partial_{j}\partial_{i}h^{k}_{k}+\partial_{j}\partial^{k}h_{ik}\right),\\  R_{ij}^{\hspace{2mm}ij}&=\partial_{i}\partial_{j}h^{ij}-\partial_{i}\partial^{i}h,
    \end{split}
\end{equation}
and 
\begin{equation}
    G^{ijkl}=\frac{1}{2}\left(\delta^{ik}\delta^{jl}+\delta^{il}\delta^{jk}\right)-\lambda\delta^{ij}\delta^{kl}.
\end{equation}

This latter expression is a linearized version of the generalized De Witt metric (\ref{Wi}); note that with $\lambda=1$ we recover the Fierz-Pauli Lagrangian. As a matter of fact, the implementation of $K_{ij}$ has been introduced in other perturbative analyses \cite{Fh,OC}, where it appears as a dynamic variable due to the presence of higher-order time derivatives, i.e., in the  Lagrangian occur time derivatives of this variable. In our case, the time derivative of $K_{ij}$ is not present. Thus, it is not a dynamic variable but a convenient way to rewrite the theory. \\ 
With this Lagrangian at hand, we will perform the canonical analysis but considering separately the cases  $\lambda\neq\frac{1}{3}$ and $\lambda=\frac{1}{3}$ since, as will be clarified below, the latter is a singular value of the theory. 
\section{Canonical analysis for $\lambda\neq\frac{1}{3}$}
We start by calculating  the canonical momenta of the action  (\ref{Lambda}), they are given by 
\begin{equation}\label{p1}
\pi^{00}=\frac{\partial\mathcal{L}}{\partial \dot{h}_{00}}=0,
\end{equation}
\begin{equation}\label{p2}
\pi^{0i}=\frac{\partial\mathcal{L}}{\partial \dot{h}_{0i}}=0,
\end{equation}
\begin{equation}\label{m1}
\hspace{1.3cm}\pi^{ij}=\frac{\partial\mathcal{L}}{\partial \dot{h}_{ij}}=G^{ijkl}K_{kl}.
\end{equation}
For constructing  the canonical Hamiltonian we need an expression for the velocity $\dot{h}_{ij}$ in terms of the canonical variables. We achieve this by considering  (\ref{m1}) and its trace
\begin{equation}
    K_{ij}=\pi_{ij}+\frac{\lambda}{1-3\lambda}\delta_{ij}\pi=\mathcal{G}_{ijkl}\pi^{kl},
\end{equation}
and reinserting the definition (\ref{k}) of $K_{ij}$ we find 
\begin{equation}\label{hvel}\dot{h}_{ij}=2\mathcal{G}_{ijkl}\pi^{kl}+\partial_{i}h_{j0}+\partial_{j}h_{i0},
\end{equation}
where $\pi=\delta_{ij}\pi^{ij}$ and $\mathcal{G}_{ijkl}=\frac{1}{2}\left(\delta_{ik}\delta_{jl}+\delta_{il}\delta_{jk}\right)+\frac{\lambda}{1-3\lambda}\delta_{ij}\delta_{kl}$ is the inverse of the generalized De Witt metric: $G^{ijkl}\mathcal{G}_{klpq}=\frac{1}{2}\left(\delta^{i}_{p}\delta^{j}_{q}+\delta^{i}_{q}\delta^{j}_{p}\right)$. It should be noted that $\lambda=\frac{1}{3}$ is a singular value of $\mathcal{G}$, therefore the treatment of this case will be  discussed in the following section.
 Now, by using (\ref{hvel}), we arrive to the canonical Hamiltonian,  it is given by
\begin{equation}
    \begin{split}
        \mathcal{H}=&\pi^{ij}\dot{h}_{ij}-\mathcal{L}=\mathcal{G}_{ijkl}\pi^{kl}\pi^{ij}-2h_{j0}\partial_{i}\pi^{ij}+\frac{1}{2}h^{00}R_{ij}^{\hspace{2mm}ij}+\frac{1}{2}h^{ij}\left(R_{ikj}^{\hspace{3.5mm}k}-\frac{1}{2}\delta_{ij}R_{lm}^{\hspace{2mm}lm}\right).
    \end{split}
\end{equation}
We identify the set of primary constraints, which are given by (\ref{p1}) and (\ref{p2})
\begin{eqnarray}
\label{pri1}
  \nonumber 
      \phi&:&\pi^{00}\approx0, \nonumber \\
      \phi^{i}&: &\pi^{0i}\approx0.
\end{eqnarray}
Thus, the primary Hamiltonian takes the form 
\begin{equation}
    \mathcal{H}'=\mathcal{G}_{ijkl}\pi^{kl}\pi^{ij}-2h_{j0}\partial_{i}\pi^{ij}+\frac{1}{2}h^{00}R_{ij}^{\hspace{2mm}ij}+\frac{1}{2}h^{ij}\left(R_{ikj}^{\hspace{3.5mm}k}-\frac{1}{2}\delta_{ij}R_{lm}^{\hspace{2mm}lm}\right)+u\phi+u_{i}\phi^{i},
\end{equation}
where $u$ and $u_{i}$ are the Lagrange multipliers enforcing the primary constraints. Then, by using the fundamental  Poisson-brackets relations 
\begin{equation}
    \left\lbrace h_{ij}(x),\pi^{kl}(y)\right\rbrace=\frac{1}{2}\left(\delta_{i}^{k}\delta_{j}^{l}+\delta_{i}^{l}\delta_{j}^{k}\right)\delta^{3}(x-y), 
\end{equation} 
and from consistency on the primary constraints i.e., that they are preserved in time,   we obtain two secondary constraints given by 
\begin{eqnarray}\label{pri2}
      \psi&:& =\dot{\phi}=\bigl\{ \phi,\int d^{3}x\hspace{1mm}\mathcal{H}' \bigr\}=R_{ij}^{\hspace{2mm}ij}\approx0, \\
      \psi^{i}&: &= \dot{\phi}^i=\bigl\{ \phi^{i},\int d^{3}x\hspace{1mm}\mathcal{H}' \bigr\}=\partial_{j}\pi^{ji}\approx0, 
      \label{pri22}
\end{eqnarray}
we can observe that the constraints (\ref{pri2}) and (\ref{pri22}) are  equivalent to the  so-called hamiltonian and momentum constraints  reported in \cite{park1}.  The process continues by applying the same criteria on these secondary  constraints. From consistency of $\psi$   we obtain a tertiary constraint
\begin{equation}\label{21s}
    \theta:\left(\frac{\lambda-1}{1-3\lambda}\right)\nabla^{2}\pi+\partial_{i}\partial_{j}\pi^{ij}\approx0,
\end{equation}
and  from the time evolution of the above expression the following constraint arise 
\begin{equation}\label{22s}
\gamma=\dot{\theta}:\left(\frac{\lambda-1}{1-3\lambda}\right)\left(\nabla^{2}\nabla^{2}h^{00}+\frac{1}{2}\nabla^{2}R_{ij}^{\hspace{2mm}ij}\right)\approx0.
\end{equation}
It is worth commenting that the constraints (\ref{21s}) and (\ref{22s}) are not reported in the literature. In this sense, our approach extends those results.  In this manner,  the generation of constraints ends, the attempt to obtain more constraints only leads to relations involving Lagrange multipliers $u$ and $u_{i}$. We have obtained a set of 10 constraints
\begin{eqnarray}
  \nonumber \phi&:&\pi^{00}\approx0, \nonumber \\
      \phi^{i}&: &\pi^{0i}\approx0,\nonumber\\
      \psi&:&R_{ij}^{\hspace{2mm}ij}\approx0, \nonumber \\
      \psi^{i}&: &\partial_{j}\pi^{ji}\approx0,\nonumber\\
       \theta&:&\left(\frac{\lambda-1}{1-3\lambda}\right)\nabla^{2}\pi\approx0,\nonumber\\
       \gamma&:&\left(\frac{\lambda-1}{1-3\lambda}\right)\left(\nabla^{2}\nabla^{2}h^{00}+\frac{1}{2}\nabla^{2}R_{ij}^{\hspace{2mm}ij}\right)\approx0,
\end{eqnarray}
which we will now proceed to classify into first-class and second-class constraints. For this purpose let us first look at the calculation of  the Poisson brackets  between the constraints
\begin{eqnarray} \nonumber\left\lbrace\gamma,\phi\right\rbrace&=&\left(\frac{\lambda-1}{1-3\lambda}\right)\nabla^{2}\nabla^{2}\delta^{3}(x-y), \nonumber \\
      \left\lbrace\psi,\theta\right\rbrace&:=&-2\left(\frac{\lambda-1}{1-3\lambda}\right)\nabla^{2}\nabla^{2}\delta^{3}(x-y),\nonumber\\
     \left\lbrace\gamma,\theta\right\rbrace&=&-\left(\frac{\lambda-1}{1-3\lambda}\right)^{2}\nabla^{2}\nabla^{2}\nabla^{2}\delta^{3}(x-y),
\end{eqnarray}
while the Poisson brackets generated by $\phi^{i}$ and $\psi^{i}$ with all other constraints vanish. The constraints whose Poisson brackets vanish with all the set of constraints are the first-class constraints and generate gauge transformations \cite{14}. Then, we identify the following  6 first-class constraints 
\begin{eqnarray}
  \nonumber 
\Gamma_{1}^{i}&: &\pi^{0i}\approx0,\nonumber\\
\Gamma_{2}^{i}&: &\partial_{j}\pi^{ji}\approx0.
\end{eqnarray}
In the opposite case, the constraints that presents  at least one Poisson brackets that do not vanish are called  second-class constraints. We identify  the following four  constraints of this kind
\begin{eqnarray}\label{sec-class}
  \nonumber  
\chi_{1}&:&R_{ij}^{\hspace{2mm}ij}\approx0, \nonumber \\ \chi_{2}&:&\left(\frac{\lambda-1}{1-3\lambda}\right)\nabla^{2}\pi\approx0,\nonumber\\
       \chi_{3}&:&\pi^{00}\approx0, \nonumber \\
       \chi_{4}&:&\left(\frac{\lambda-1}{1-3\lambda}\right)\nabla^{2}\nabla^{2}h^{00}\approx0.
\end{eqnarray}
In this manner, the counting of the degrees of freedom is carried out in the following form 
\begin{equation}
    DOF=\frac{1}{2}(\mathrm{canonical\hspace{1mm}var.}-2(\mathrm{first\hspace{1mm}class\hspace{1mm}c.})-\mathrm{second\hspace{1mm}class\hspace{1mm}c.})=\frac{1}{2}(20-4-2*6)=2. 
\end{equation}
This is consistent with that reported in  \cite{c2}, that the extra mode is excited in perturbative analyses only on time-dependent and spatially non-homogeneous backgrounds. It is worth commenting that 
in the special case $\lambda=1$, then $\dot{\psi}=0$, and the only remaining constraints are (\ref{pri1}) and (\ref{pri2}) which are first-class. As mentioned before, this corresponds to $GR$, where  the constraints are  first-class and again  $DOF=\frac{1}{2}(20-2*8)=2$. \\
Furthermore, second-class constraints are not gauge generators. The proper way to handle them was introduced by Dirac. In fact, they are removed by introducing the Dirac brackets
\begin{equation}
    \left\lbrace A, B\right\rbrace_{D}= \left\lbrace A, B\right\rbrace-\int dudv\left\lbrace A, \chi_{a}(u)\right\rbrace C^{ab}\left\lbrace \chi_{b}(v), B\right\rbrace,
\end{equation}
where $C^{ab}$ is the inverse of $C_{ab}=\left\lbrace \chi_{a}, \chi_{b}\right\rbrace$.  In this way, we remove the second-class constraints, and all the dynamical equations of the theory are expressed in terms of these brackets. The second-class constraints are relevant in either the construction of the extended Hamiltonian or the identification of Lagrange multipliers; the number of second-class constraints in any gauge theory indicates the number of Lagrange multipliers that can be identified \cite{14}.  \\
In this way, considering (\ref{sec-class}) we find 
\begin{equation}\label{matrix1}
\makeatletter\setlength\BA@colsep{7pt}\makeatother
\arraycolsep=1.4pt\def\arraystretch{0.85}
   C_{\alpha\beta}=
    \begin{blockarray}{ccccc}
        & \chi_{1} & \chi_{2} & \chi_{3} & \chi_{4} \\
      \begin{block}{c(cccc)}
        \chi_{1} & 0 & -2\beta\nabla^{2}\nabla^{2} & 0 & 0  \\
        \chi_{2} & 2\beta\nabla^{2}\nabla^{2} & 0 & 0 & 0  \\
        \chi_{3} &0 & 0 & 0 & -\beta\nabla^{2}\nabla^{2} \\
        \chi_{4} & 0 & 0 & \beta\nabla^{2}\nabla^{2} & 0  \\
      \end{block}
    \end{blockarray}\hspace{1mm}\delta^{3}(x-y),
\end{equation}
where $\beta=\frac{\lambda-1}{1-3\lambda}$. Its inverse is given by 
\begin{equation}\label{matrix2}
\makeatletter\setlength\BA@colsep{7pt}\makeatother
\arraycolsep=1.4pt\def\arraystretch{0.85}
   C^{\alpha\beta}=
    \begin{blockarray}{ccccc}
        & \chi_{1} & \chi_{2} & \chi_{3} & \chi_{4} \\
      \begin{block}{c(cccc)}
        \chi_{1} &  0 & \frac{1}{2} & 0 & 0  \\
        \chi_{2} &- \frac{1}{2} & 0 & 0 & 0  \\
        \chi_{3} & 0 & 0 & 0 &1 \\
        \chi_{4} & 0 & 0 & -1 & 0  \\
      \end{block}
    \end{blockarray}\hspace{1mm}\frac{1}{\beta\nabla^{2}\nabla^{2}}\delta^{3}(x-y).
\end{equation}
Thus, the non vanishing Dirac's  brackets are 
\begin{equation}
    \left\lbrace h_{ij},\pi^{lm}\right\rbrace_{D}=\frac{1}{2}\left(\delta_{i}^{l}\delta_{j}^{m}+\delta_{i}^{m}\delta_{l}^{k}\right)\delta^{3}(x-y)+\frac{1}{2\nabla^{2}}\delta_{ij}\left(\partial^{l}\partial^{m}-\delta^{lm}\nabla^{2}\right)\delta^{3}(x-y).
\end{equation}
Let us  calculate  the Dirac brackets between the constraints and the canonical Hamiltonian of this model. For the second-class constraints we have 
\begin{equation}
\begin{split}
 \left\lbrace \chi^{1},\mathcal{H}\right\rbrace_{D}&=0,
 \\
 \left\lbrace \chi^{2},\mathcal{H}\right\rbrace_{D}&=0,
    \\
     \left\lbrace \chi^{3},\mathcal{H}\right\rbrace_{D}&=0,\\
     \left\lbrace \chi^{4},\mathcal{H}\right\rbrace_{D}&=0,
\end{split}
\end{equation}
and its algebra with the first-class constraints is given by
\begin{eqnarray}
\nonumber
    \left\lbrace \Gamma_{1}^{i}, \mathcal{H}\right\rbrace_{D}&=&\Gamma_{2}^{i},\nonumber \\
    \left\lbrace \Gamma_{2}^{i}, \mathcal{H}\right\rbrace_{D}&=&0.
\end{eqnarray}
Thus the algebra closes and the Hamiltonian is of first-class.

\section{Canonical analysis for $\lambda=\frac{1}{3}$}

With this value of $\lambda$ let us consider the particular form  of the generalized De Witt metric
\begin{equation}
    \hat{G}^{ijkl}=\frac{1}{2}\left(\delta^{ik}\delta^{jl}+\delta^{il}\delta^{jk}\right)-\frac{1}{3}\delta^{ij}\delta^{kl}.
\end{equation}
Thus, the Lagrangian is written as
\begin{equation}
    \mathcal{L}=\hat{G}^{ijkl}K_{ij}K_{kl}-\frac{1}{2}h^{00}R_{ij}^{\hspace{2mm}ij}-\frac{1}{2}h^{ij}\left(R_{ikj}^{\hspace{3.5mm}k}-\frac{1}{2}\delta_{ij}R_{lm}^{\hspace{2mm}lm}\right),
\end{equation}
now the expressions for the canonical momenta are 
\begin{equation}
    \pi^{00}=\frac{\partial\mathcal{L}}{\partial \dot{h}_{00}}=0,
\end{equation}
\begin{equation}
    \pi^{0i}=\frac{\partial\mathcal{L}}{\partial \dot{h}_{0i}}=0,
\end{equation}
\begin{equation}\label{m2}
    \hspace{1.3cm}\pi^{ij}=\frac{\partial\mathcal{L}}{\partial \dot{h}_{ij}}=\hat{G}^{ijkl}K_{kl}.
\end{equation}
In order to construct the canonical Hamiltonian,  we observe   from (\ref{m2}) that  $\delta_{ij}\pi^{ij}=\pi=0$ and 
\begin{equation}
    \pi^{ij}K_{ij}=K_{ij}K^{ij}-\frac{1}{3}K^{2}=\pi^{ij}\pi_{ij}
\end{equation}
 Thus, the canonical Hamiltonian is given by 
\begin{equation}
    \begin{split}
        \mathcal{H}=&\pi^{ij}\dot{h}_{ij}-\mathcal{L}=2\pi^{ij}K_{ij}+2\pi^{ij}\partial_{i}h_{0j}-\mathcal{L}\\=&
        \pi^{ij}\pi_{ij}-2\partial_{i}\pi^{ij}h_{0j} +\frac{1}{2}h^{00}R_{ij}^{\hspace{2mm}ij}+\frac{1}{2}h^{ij}\left(R_{ikj}^{\hspace{3.5mm}k}-\frac{1}{2}\delta_{ij}R_{lm}^{\hspace{2mm}lm}\right).
    \end{split}
\end{equation}
In this case,  the primary constraints are identified as  
\begin{eqnarray}
\label{conpri}
  \nonumber 
      \phi&:&\pi^{00}\approx0, \nonumber \\
      \phi^{i}&: &\pi^{0i}\approx0,\nonumber\\
      \xi&:&\pi\approx0,
\end{eqnarray}
at this point we would comment that the constraint $\pi\approx0$ is not reported in \cite{park2}, however, it is found in the nonperturbative analysis developed in \cite{bellorin}. Hence, the primary Hamiltonian takes the form 
\begin{equation}
        \mathcal{H}'=\pi^{ij}\pi_{ij}-2\partial_{i}\pi^{ij}h_{0j} +\frac{1}{2}h^{00}R_{ij}^{\hspace{2mm}ij}+\frac{1}{2}h^{ij}\left(R_{ikj}^{\hspace{3.5mm}k}-\frac{1}{2}\delta_{ij}R_{lm}^{\hspace{2mm}lm}\right)+u\phi+u_{i}\phi^{i}+v\xi.
\end{equation}
From consistency on the primary constraints we obtain the following five secondary constraints 
\begin{eqnarray}
  \label{433}
      \psi&:&R_{ij}^{\hspace{2mm}ij}\approx0, \\
      \psi^{i}&: &\partial_{j}\pi^{ji}\approx0,\label{443} \\
      \gamma&:&\nabla^{2}h^{00}+\frac{1}{2}R_{ij}^{\hspace{2mm}ij}\approx0 \label{453}, 
\end{eqnarray}
again, the constraints (\ref{433}) and (\ref{443}) are the so-called hamiltonian and momentum constraints respectively, they were reported in \cite{park2}, however, (\ref{453})  was not identified. Then, the evolution of these expressions  results in the following  relations between Lagrange multipliers
\begin{eqnarray}
  \nonumber  
      \dot{\psi}&:&\partial_{i}\partial_{j}\pi^{ij}-\nabla^{2}\pi-\nabla^{2}v\approx0, \nonumber \\
      \dot{\gamma}&:&\nabla^{2}u\approx0.
\end{eqnarray}
Therefore, the complete set of constraints is 
\begin{eqnarray}
\label{conpri}
  \nonumber 
      \phi&:&\pi^{00}\approx0, \nonumber \\
      \phi^{i}&: &\pi^{0i}\approx0,\nonumber\\
      \xi&:&\pi\approx0,\nonumber\\
       \psi&:&R_{ij}^{\hspace{2mm}ij}\approx0, \nonumber \\
      \psi^{i}&: &\partial_{j}\pi^{ji}\approx0,\nonumber\\
      \gamma&:&\nabla^{2}h^{00}+\frac{1}{2}R_{ij}^{\hspace{2mm}ij}\approx0,
\end{eqnarray}
and the nonzero Poisson brackets between them are
\begin{equation} 
\begin{split}
    \left\lbrace\gamma,\phi\right\rbrace&=\nabla^{2}\delta^{3}(x-y),\\
    \left\lbrace\psi,\xi\right\rbrace&=-2\nabla^{2}\delta^{3}(x-y),\\
    \left\lbrace\gamma,\xi\right\rbrace&=-\nabla^{2}\delta^{3}(x-y).
\end{split}
\end{equation}
With this result we can perform the classification of constraints,  thus, we obtain  6 first-class constraints given by 
\begin{eqnarray}\label{fc1/3}
  \nonumber 
      \Gamma_{1}^{i}&: &\pi^{0i}\approx0,\nonumber\\
      \Gamma_{2}^{i}&: &\partial_{j}\pi^{ji}\approx0,
\end{eqnarray}
and the following 4 second-class constraints 
\begin{eqnarray}\label{sc}
  \nonumber  
      \chi_{1}&:&R_{ij}^{\hspace{2mm}ij}\approx0, \nonumber \\
       \chi_{2}&:&\pi\approx0,\nonumber\\
       \chi_{3}&:&\pi^{00}\approx0, \nonumber \\
       \chi_{4}&:&\nabla^{2}h^{00}\approx0.
\end{eqnarray}
Such as  the value  of $\lambda\neq \frac{1}{3}$, the counting of the degrees of freedom yields two. We will also construct the Dirac brackets, for which, as we commented above, it is necessary to calculate the matrix between the second-class constraints and their inverse. These are given by 
\begin{equation}\label{matrix3}
\makeatletter\setlength\BA@colsep{7pt}\makeatother
\arraycolsep=1.4pt\def\arraystretch{0.85}
   C_{\alpha\beta}=
    \begin{blockarray}{ccccc}
        & \chi_{1} & \chi_{2} & \chi_{3} & \chi_{4} \\
      \begin{block}{c(cccc)}
        \chi_{1} & 0 & -2\nabla^{2} & 0 & 0  \\
        \chi_{2} & 2\nabla^{2} & 0 & 0 & 0  \\
        \chi_{3} &0 & 0 & 0 & -\nabla^{2} \\
        \chi_{4} & 0 & 0 &\nabla^{2} & 0  \\
      \end{block}
    \end{blockarray}\hspace{1mm}\delta^{3}(x-y),
\end{equation}
and its inverse takes the form 
\begin{equation}\label{matrix4}
\makeatletter\setlength\BA@colsep{7pt}\makeatother
\arraycolsep=1.4pt\def\arraystretch{0.85}
   C^{\alpha\beta}=
    \begin{blockarray}{ccccc}
        & \chi_{1} & \chi_{2} & \chi_{3} & \chi_{4} \\
      \begin{block}{c(cccc)}
        \chi_{1} &  0 & \frac{1}{2} & 0 & 0  \\
        \chi_{2} &- \frac{1}{2} & 0 & 0 & 0  \\
        \chi_{3} & 0 & 0 & 0 &1 \\
        \chi_{4} & 0 & 0 & -1 & 0  \\
      \end{block}
    \end{blockarray}\hspace{1mm}\frac{1}{\nabla^{2}}\delta^{3}(x-y).
\end{equation}
In this manner, the nonzero Dirac's brackets are given by 
\begin{equation}
    \left\lbrace h_{ij},\pi^{lm}\right\rbrace_{D}=\frac{1}{2}\left(\delta_{i}^{l}\delta_{j}^{m}+\delta_{i}^{m}\delta_{l}^{k}\right)\delta^{3}(x-y)+\frac{1}{2\nabla^{2}}\delta_{ij}\left(\partial^{l}\partial^{m}-\delta^{lm}\nabla^{2}\right)\delta^{3}(x-y),
\end{equation}
where we obtain the same brackets of the previous section,  and  the Hamiltonian for this case is also of  first-class. \\
The equivalence between $\lambda R$ gravity and $GR$ in this perturbative approach becomes further evident by fixing the gauge. For this purpose let us consider  the Coulomb gauge $\partial_{i}h^{ij}\approx0$ together with $h^{0i}\approx0$ which are agree with the first-class constraints obtained above. Therefore, the set of second-class constraints becomes 
\begin{eqnarray}\label{scgf}
  \nonumber  
      \chi_{1}&:&h_{i}^{i}\approx0, \nonumber \\
       \chi_{2}&:&\pi\approx0,\nonumber\\
       \chi_{3}&:&\pi^{00}\approx0, \nonumber \\
       \chi_{4}&:&\nabla^{2}h^{00}\approx0,\nonumber \\
     \chi_{5}&: &\pi^{0i}\approx0,\nonumber\\
     \chi_{6}&:&h^{0i}\approx0,\nonumber\\
     \chi_{7}&:&\partial_{j}\pi^{ji}\approx0,\nonumber\\
     \chi_{8}&:&\partial_{j}h^{ji}\approx0.
\end{eqnarray}
After a long algebraic work, we obtain that the non-vanishing Dirac brackets that follow from this set of second-class constraints are 
\begin{equation}\label{secondfull}
\begin{split}
     \left\lbrace h_{ij},\pi^{lm}\right\rbrace_{D}&=\frac{1}{2}\left(\delta_{i}^{l}\delta_{j}^{m}+\delta_{i}^{m}\delta_{j}^{l}\right)\delta^{3}(x-y)-\frac{1}{2\nabla^{2}}\left(\delta_{i}^{m}\partial_{j}\partial^{l}+\delta_{i}^{l}\partial_{j}+\delta_{j}^{m}\partial_{i}\partial^{l}+\delta_{j}^{l}\partial_{i}\partial^{m}\right)\delta^{3}(x-y)\\
     &-\frac{1}{2}\delta_{ij}\delta^{lm}\delta^{3}(x-y)+\frac{1}{2\nabla^{2}}\left(\delta_{ij}\partial^{l}\partial^{m}+\delta^{lm}\partial_{i}\partial_{j}\right)\delta^{3}(x-y)+\frac{1}{2}\frac{\partial_{i}\partial_{j}\partial^{l}\partial^{m}}{\nabla^{4}}\delta^{3}(x-y).
\end{split}
\end{equation} 
These brackets correspond to those found in  \cite{Bar, Mel} for linearized gravity, where the gauge is fixed completely by using the conditions  $\pi\approx0$, $\partial_{i}h^{ij}\approx0$, $h_{0i}\approx0$ and $h_{00}\approx0$. Indeed, the set of second-class constraints obtained in this way is the same in both models. 
Furthermore, from our results it is possible to obtain a remarkable result found   in the full  model (\ref{full}) \cite{bellorin}. In fact, it arises from the  equation of motion for $h_{ij}$
\begin{equation}
    \dot{h}_{ij}=\left\lbrace h_{ij},\mathcal{H}\right\rbrace_D=2\pi_{ij}+\partial_{i}h_{0j}+\partial_{j}h_{0i},
\end{equation}
then $\delta^{ij}\dot{h}_{ij}=\dot{h}_{i}^{i}=2\pi+2\partial_{i}h_{0}^{i}$ and by using the second-class constraint $\pi\approx0$ we obtain the following  expression
\begin{equation}
    K=\frac{1}{2}\left(\dot{h}_{i}^{i}-2\partial_{i}h_{0}^{i}\right)\approx0,
\end{equation}
which is of  second-class as it can be seen by using  (\ref{secondfull}). As a consequence, in the constrained phase space the term $\lambda K^{2}$  is no relevant, so that the constant $\lambda$  no longer promotes a distinction between $GR$ and $\lambda R$ gravity. 
\section{Conclusions}
From the perturbative point of view,  the complete set of first-class and second-class constraints for the  Hořava-like theory, with values of  $\lambda=\frac{1}{3}$ and $\lambda\neq\frac{1}{3}$ were obtained. As far as we know,  these results have not been reported in the literature. It is important to remark that our purpose is to perform the canonical analysis more economically instead of working with the perturbative  $ADM$  variables as it is usually done.   In this regard, in the analysis reported in \cite{park1, park2}, the complete set of  constraints  and the Dirac brackets were not reported. Furthermore,  in these works, the identification of the constraint $\pi \approx 0$  is not evident for $\lambda\neq\frac{1}{3}$, so it is concluded that there is an extra degree of freedom. Although it is argued that this degree of freedom is not physical, then it is removed by fixing the gauge. In this respect, in our approach, no extra conditions are used; the identification of the constraints and the counting of the physical degrees of freedom were easily performed. Moreover, we have shown that at the perturbative level, the Dirac brackets for every value of $\lambda$ coincide; then,  the theory is equivalent to $GR$  independent of $\lambda$, this agrees with the results obtained in \cite{bellorin} where a non-perturbative approach was reported. \\
It is worth mentioning that part of the discussion about the inconsistency of the Ho\v{r}ava theory was based on the assumption that theory is compatible with $GR$ at large distances only when $\lambda\rightarrow 1$, i.e., when the full diffeomorphism group is restored. Adopting this stance, the analysis  performed in \cite{c1} provided partial conclusions about the strong coupling in the different possible versions of the theory. However, as mentioned in the introduction,
the $\lambda R$ model supports the compatibility of Ho\v{r}ava gravity with $GR$  at large distances.
On the other hand, in \cite{c2}, it was shown that the extra mode present at short distances is of an odd nature; that is, it propagates itself with a first-order time derivative.
With the aim of curing the oddness of the extra mode, it was noticed in \cite{blasf} that, once the principle of detailed balance is discarded, the nonprojectable Ho\v{r}ava action admits a large class of terms and the extra mode becomes even (propagates with a second-order time derivative) in the complete theory.
Therefore, by using our approach  will be interesting to develop  the study of  the closeness  of the constraint algebra and renormalizability,  in the possible extensions of the nonprojectable Ho\v{r}ava theory \cite{bellorin2, bellorin3,bellorin4}, however, this will be the subject of forthcoming works.

\section{Appendix A}
This appendix is added to deriving the Fierz-Pauli Lagrangian (\ref{FP}). To this end, we start with the $EH$ action
\begin{equation}
   S_{EH}=\int d^{4}x\hspace{2mm}\sqrt{-g}R.
\end{equation}
The linearization procedure begins by considering the second-order approximation of the Ricci scalar
\begin{equation}
\begin{split}
     R&=\eta^{\mu\nu} R^{(lin)}_{\mu\nu}-\epsilon h^{\mu\nu}R^{(lin)}_{\mu\nu}\\&
     =\epsilon\left(\partial_{\mu}\partial_{\nu}h^{\mu\nu}-\Box h\right)-\frac{1}{2}\epsilon^{2} h^{\mu\nu}\left(\partial_{\alpha}\partial_{\mu}h_{\nu}^{\alpha}+\partial_{\alpha}\partial_{\nu}h_{\mu}^{\alpha}-\partial_{\mu}\partial_{\nu}h-\Box h_{\mu\nu}\right),
\end{split}
\end{equation}
where $R^{(lin)}_{\mu\nu}$ is the first-order Ricci tensor and the perturbation is $g_{\mu\nu}=\eta_{\mu\nu}+\epsilon h_{\mu\nu}$.
Then 

\begin{equation}
    \begin{split}
        S_{EH}&=\int d^{4}x\sqrt{-\eta}(1+\epsilon\frac{1}{2}h)\left[\epsilon\left(\partial_{\mu}\partial_{\nu}h^{\mu\nu}-\Box h\right)-\frac{1}{2}\epsilon^{2} h^{\mu\nu}\left(\partial_{\alpha}\partial_{\mu}h_{\nu}^{\alpha}+\partial_{\alpha}\partial_{\nu}h_{\mu}^{\alpha}-\partial_{\mu}\partial_{\nu}h-\Box h_{\mu\nu}\right)\right]\\&
        =\epsilon^{2}\int d^{4}x\sqrt{-\eta}\left[-\frac{1}{2} h^{\mu\nu}\left(\partial_{\alpha}\partial_{\mu}h_{\nu}^{\alpha}+\partial_{\alpha}\partial_{\nu}h_{\mu}^{\alpha}-\partial_{\mu}\partial_{\nu}h-\Box h_{\mu\nu}\right)+\frac{1}{2}h\left(\partial_{\mu}\partial_{\nu}h^{\mu\nu}-\Box h\right)\right]\\&
        =\epsilon^{2}\int d^{4}x\sqrt{-\eta}\left[  \partial_{\mu}h^{\mu\nu}\partial_{\alpha}h_{\nu}^{\alpha}-\frac{1}{2}\partial^{\alpha} h^{\mu\nu}\partial_{\alpha}h_{\mu\nu} -\partial_{\nu}h\partial_{\mu}h^{\mu\nu}+\frac{1}{2}\partial_{\alpha}h \partial^{\alpha}h\right].
    \end{split}
\end{equation}
The first-order part is a total divergence and we have kept only the terms up to second-order in $\epsilon$. In this way  we obtain the next linearized Lagrangian.
\begin{equation}
\mathcal{L}_{FP}=\frac{1}{2}\partial^{\alpha}h^{\mu\nu}\partial_{\mu}h_{\nu\alpha}-\frac{1}{4}\partial^{\alpha} h^{\mu\nu}\partial_{\alpha}h_{\mu\nu} -\frac{1}{2}\partial_{\nu}h\partial_{\mu}h^{\mu\nu}+\frac{1}{4}\partial_{\alpha}h \partial^{\alpha}h.
\end{equation}
This expression is known as the
Fierz-Pauli Lagrangian for massless particles of spin two and propagates two degrees of freedom. Its 3+1 form is obtained by using $\eta_{\mu\nu}=\mathrm{diag}(-1,1,1,1)$ and the decomposition of each term considering separately the temporal and spatial components. For instance, the first term takes the form 
\begin{equation}
\begin{split}
\partial^{\alpha}h^{\mu\nu}\partial_{\mu}h_{\nu\alpha}&=\partial^{0}h^{00}\partial_{0}h_{00}+\partial^{0}h^{0i}\partial_{0}h_{i0}+2\partial^{i}h^{00}\partial_{0}h_{i0}+2\partial^{0}h^{ij}\partial_{i}h_{0j}+\partial^{i}h^{j0}\partial_{j}h_{i0}+\partial^{i}h^{jk}\partial_{j}h_{ik}\\
     &=-\dot{h}^{00}\dot{h}_{00}-\dot{h}^{0i}\dot{h}_{0i}+2\partial^{i}h^{00}\dot{h}_{i0}-2\dot{h}^{ij}\partial_{i}h_{0j}+\partial^{i}h^{j0}\partial_{j}h_{i0}+\partial^{i}h^{jk}\partial_{j}h_{ik}.
\end{split}
\end{equation}
The decomposition of all the terms is condensed in the following expression 
\begin{equation}\label{FP2}
    \begin{split}
        \mathcal{L}_{FP}=&\frac{1}{4}\Dot{h}_{ij}\Dot{h}^{ij}-\Dot{h}^{ij}\partial_{i}h_{0j}-\dot{h}_{j}^{j}\partial_{i}h^{0i}-\frac{1}{4}(\dot{h}_{i}^{i})^{2}-\frac{1}{2}\partial_{i}h_{0j}\partial^{i}h^{0j}+\frac{1}{2}\partial^{i}h^{j0}\partial_{j}h_{i0}+\frac{1}{2}\partial_{i}h_{00}\partial_{j}h^{ij}\\&\hspace{4mm}-\frac{1}{2}\partial_{i}h_{k}^{k}\partial_{j}h^{ij}-\frac{1}{2}\partial_{i}h_{00}\partial^{i}h_{k}^{k}+\frac{1}{4}\partial_{i}h_{j}^{j}\partial^{i}h_{k}^{k}+\frac{1}{2}\partial^{i}h^{jk}\partial_{j}h_{ik}-\frac{1}{4}\partial_{i}h_{jk}\partial^{i}h^{jk}.
    \end{split}
\end{equation}
 From the canonical point of view, the  classification of constraints as well as the construction of the Dirac brackets of this theory,  is worked on in \cite{Bar}. Moreover, from the symplectic point of view, this theory was studied in \cite{Mel}.




\end{document}